\documentclass[10pt]{iopart}

\usepackage{color}

\usepackage{hyperref}
\usepackage{xspace}
\expandafter\let\csname equation*\endcsname\relax
\expandafter\let\csname endequation*\endcsname\relax
\usepackage{amsmath}

\usepackage{amsthm}
\usepackage{amssymb}
\usepackage{amsfonts}
\usepackage{dsfont}
\usepackage{accents}
\usepackage{mathtools}
\usepackage{tabularx} 
\usepackage{caption}
\usepackage{multirow}
\usepackage{booktabs}
\usepackage{graphicx}
\usepackage{subfigure}
\usepackage{tikz,pgfplots}
\pgfplotsset{compat=1.10}
\usepgfplotslibrary{groupplots}
\usetikzlibrary{intersections,calc,arrows,matrix,spy}
\usepackage{color}
\definecolor{darkred}{rgb}{0.6,0,0}
\definecolor{darkgreen}{rgb}{0,0.5,0}
\definecolor{darkblue}{rgb}{0,0,0.5}
\definecolor{lightgrey}{rgb}{0.96,0.96,0.96}
\definecolor{SkyBlue}{rgb}{0.53, 0.81, 0.92}
\definecolor{lilas}{RGB}{170,55,241}
\pgfplotsset{compat=1.5.1}
\usepackage{textcomp}
\usetikzlibrary{shapes,arrows}
\tikzset{%
  abstractClass/.style    = {draw, thick, rectangle, minimum height = 2.5em,
    minimum width = 6em,fill=blue!20,rounded corners},
  concreteClass/.style    = {draw, thick, rectangle, minimum height = 2.5em,
    minimum width = 6em,fill=gray!20,rounded corners},
  blank/.style    = {rectangle, minimum height = 2.5em,
    minimum width = 6em,rounded corners},
    cross/.style={path picture={ \draw[black] (path picture bounding box.west) -- (path picture bounding box.east) (path picture bounding box.south) -- (path picture bounding box.north);}},
  sum/.style      = {draw, circle, cross,node distance = 2cm,minimum width=1 cm}, 
  prod/.style      = {draw, circle,node distance = 2cm,minimum width=1 cm}, 
  input/.style    = {coordinate}, 
  output/.style   = {coordinate} 
}
\usepackage{algorithm}
\usepackage{algorithmic}
\usepackage[squaren,Gray]{SIunits}
\usepackage{listings}
\usepackage{lstlinebgrd}
\newcommand{\argmin}[2]{\mathrm{arg}\,\underset{#1}{  \mathrm{min}} \; #2}  
\def\R{\mathbb{R}}          									 	          

\newcommand{\Dd}{\mathrm{\mathbf{D}}}

\newcommand{\Hd}{{\mathbf{H}}}
\newcommand{\Id}{\mathrm{\mathbf{I}}}
\newcommand{\Jd}{\mathrm{\mathbf{J}}}
\newcommand{\Ld}{{\mathbf{L}}}

\newcommand{\Sd}{{\mathbf{S}}}
\newcommand{\Td}{\mathrm{\mathbf{T}}}

\newcommand{\fd}{{\mathbf{f}}}
\newcommand{\gd}{{\mathbf{g}}}

\newcommand{\nd}{{\mathbf{n}}}

\newcommand{\ud}{\mathrm{\mathbf{u}}}
\newcommand{\vd}{\mathrm{\mathbf{v}}}

\newcommand{\zd}{\mathrm{\mathbf{z}}}

\def\nab{\mbox{{\boldmath{$\nabla$}}}}

\newcommand{\disp}{\displaystyle}                                           
\newcommand{\ie}{\textit{i.e., }}                                            
\newcommand{\eg}{\textit{e.g., }}                                            
  
\def\map{{\tt Map}\xspace}        
\def\cost{{\tt Cost}\xspace}    
\def\linop{{\tt LinOp}\xspace}    
\def\opti{{\tt Opti}\xspace}   
\def\globalbioim{{\tt GlobalBioIm}\xspace}
\newcommand{\meth}[1]{{\tt  #1}\xspace}

\theoremstyle{remark}

\newtheorem{example}{Example}[section]

\usepackage[framemethod=tikz]{mdframed}
\makeatletter
\mdf@do@stringoption{digressiontitle=={Digression}}
\tikzset{
  excursus arrow/.style={%
       line width=2pt,
      draw=blue!20,
      rounded corners=2ex,
      },
   excursus head/.style={
      fill=white,
      font=\bfseries\sffamily,
      text=black,
      anchor=base west,
  },
}
\mdfdefinestyle{digressionarrows}{%
   singleextra={%
            \path let \p1=(P), \p2=(O) in (\x2,\y1) coordinate (Q);
            \path let \p1=(Q), \p2=(O) in (\x1,{(\y1-\y2)/2}) coordinate (M);
            \path [excursus arrow, round cap-to]
                        ($(O)+(5em,0ex)$) -| (M) |- %
                        ($(Q)+(12em,0ex)$) .. controls +(0:16em) and +(185:0em) .. %
                        ++(23em,0ex);
            \node [excursus head] at ($(Q)+(2.5em,0ex)$) {\mdf@digressiontitle};},
   firstextra={%
            \path let \p1=(P), \p2=(O) in (\x2,\y1) coordinate (Q);
            \path [excursus arrow,-to] (O) |- %
                        ($(Q)+(12em,0ex)$) .. controls +(0:16em) and +(185:0em) .. %
                        ++(23em,0ex);
            \node [excursus head] at ($(Q)+(2.5em,0ex)$) {\mdf@digressiontitle};},
   secondextra={%
            \path let \p1=(P), \p2=(O) in (\x2,\y1) coordinate (Q);
            \path [excursus arrow,round cap-]($(O)+(5em,0ex)$) -| (Q);},
   middleextra={%
            \path let \p1=(P), \p2=(O) in (\x2,\y1) coordinate (Q);
            \path [excursus arrow](O) -- (Q);},
   middlelinewidth=2.5em,middlelinecolor=white,
   hidealllines=true,topline=true,
   innertopmargin=0.5ex,
   innerbottommargin=2.5ex,
   innerrightmargin=2pt,
   innerleftmargin=2ex,
   skipabove=0.87\baselineskip,
   skipbelow=0.62\baselineskip,
}
\newmdenv[style=digressionarrows,digressiontitle=GlobalBioIm in Practice]{inpractice}
\begin{document}

\title{Pocket Guide to Solve Inverse Problems with GlobalBioIm}

\author{Emmanuel Soubies$^1$, Ferr\'eol Soulez$^2$, Michael T. McCann$^1$, Thanh-an Pham$^1$, Laur\`ene Donati$^1$, Thomas Debarre$^1$, Daniel Sage$^1$, and Michael Unser$^1$}

\address{$^1$Biomedical Imaging Group, EPFL, Lausanne, Switzerland. \\ $^2$Univ Lyon, Univ Lyon1, Ens de Lyon, CNRS, Centre de Recherche Astrophysique de Lyon UMR5574, F-69230, Saint-Genis-Laval, France.}
\ead{michael.unser@epfl.ch}
\vspace{10pt}

%
%
%
%
%

\begin{abstract}
\globalbioim is an open-source MATLAB\textsuperscript{\textregistered} library for solving inverse problems. The library capitalizes on the strong commonalities between forward models to standardize the resolution of a wide range of imaging inverse problems. Endowed with an operator-algebra mechanism, \globalbioim allows one to easily solve inverse problems by combining elementary modules in a lego-like fashion. This user-friendly toolbox gives access to cutting-edge reconstruction algorithms, while its high modularity makes it easily extensible to new modalities and novel reconstruction methods. We expect \globalbioim to respond to the needs of imaging scientists looking for reliable and easy-to-use computational tools for solving their inverse problems. In this paper, we present in detail the structure and main features of the library. We also illustrate its flexibility with examples from multichannel deconvolution microscopy. 
\end{abstract}


\section{Introduction} 
    
    \subsection{Inverse Problems in Imaging} 
    
    Imaging is a fundamental tool for biological research, medicine, and astrophysics. Medical imaging systems are essential for modern diagnosis, while the latest generation of microscopes and telescopes provide images with unprecedented resolution. This imaging revolution is driven, in part, by the current shift towards computational imaging that sees optics and computing combine to bypass many limitations of conventional systems.
   
    These computational imaging techniques rely on the deployment of sophisticated algorithms to reconstruct a $d$-dimensional continuously defined object of interest $f \in L_2(\R^d)$ 
    from discrete measurements $\gd \in \R^M$ recorded by a given imaging system. These quantities are linked according to
   \begin{equation}
  \gd = \mathcal{H}\{f\} + \nd,
   \end{equation}
   where $\mathcal{H} : L_2(\R^d) \rightarrow \R^M$ is an operator that models the imaging system. This operator, which might be linear or not, maps the continuously defined object to discrete noiseless measurements. 
   Finally, $\nd \in \R^M$ is an error term, which is often considered to be random.  
   
   To numerically solve the inverse problem and recover $f$, it is necessary to discretize both $f$ and the operator $\mathcal{H}$. This leads to the discrete imaging model $\gd = \Hd\{\fd\} + \nd$, with $\fd \in \R^N$ and $\Hd\{\cdot \} : \R^N \rightarrow \R^M$.
   
   The classical approach to address this inverse problem and recover an estimated solution $\hat{\fd}$ consists in solving 
\begin{equation}\label{eq:OptiPb}
    \hat{\fd} = \argmin{\fd \in \R^N}{ \big( \mathcal{D}(\Hd \{\fd\}, \gd) + \lambda \mathcal{R}(\fd) \big)}.
\end{equation}
There, $\mathcal{D} : \R^M \times \R^M \rightarrow \R$ measures the discrepancy between the forward model $\Hd \{\fd\}$ and the measurements $\gd$ (\textit{i.e.}, the data fidelity), while  $\mathcal{R} : \R^N \rightarrow \R$ enforces specific regularity constraints on the solution (\eg spatial smoothness, or nonnegativity). The balance between the data fidelity and regularization terms is controlled by the scalar parameter $\lambda >0$.

\subsection{Unifying Framework for Solving Inverse Problems}

The forward models associated with most of the commonly used imaging modalities share important structural properties. This similarity is not surprising since many imaging systems are governed by the same physical principles (\eg the wave equation). We express in Table \ref{Table:modalities} the forward models of a wide range of imaging modalities in terms of a limited number of elementary constituents.

\begin{table}
		\centering
	   \caption{\label{Table:modalities} Broad class of imaging models defined as the composition of elementary operators. Here, the basic constituents include weighting, windowing, or modulation ($\mathcal{W}$), convolution ($\mathcal{C}$), Fourier transform ($\mathcal{F}$), integration ($\Sigma$), rotation ($\mathcal{R}_\theta)$, and sampling ($\mathcal{S}$).  The Radon transform (for CT and cryo-EM) is written as the composition $\Sigma \circ \mathcal{R}_{\theta}$ of a rotation and an integration. Similarly, the Laplace transform (for TIRF) is expressed as the composition $\Sigma \circ \mathcal{W}$ of a weighting (decaying exponential) and an integration. Note that these elementary operators might differ for each modality (\eg using different kernels for the convolution operators), but their construction stays identical.}
		\begin{tabular}{ll} 
			\toprule
			\toprule
			  Imaging modality &   Forward model $\mathcal{H}$  \\  \midrule
			  X-ray computed tomography (CT) & $\mathcal{S} \circ \Sigma \circ \mathcal{R}_{\theta}$\\
			  Conventional fluorescent microscopy & $\mathcal{S} \circ \mathcal{C}$ \\
			  Structured-illumination microscopy (SIM) & $\mathcal{S} \circ \mathcal{C} \circ \mathcal{W}$\\
			  Total internal reflection fluorescence  (TIRF)& $\mathcal{S} \circ \Sigma \circ \mathcal{W}$\\
			  Optical diffraction tomography (ODT, first Born)& $\mathcal{S} \circ \mathcal{C} \circ \mathcal{W}$ \\
			  Cryo-electron tomography (Cryo-EM)& $\mathcal{S} \circ \mathcal{C} \circ \Sigma \circ \mathcal{R}_{\theta}$ \\
			  Magnetic resonance imaging (MRI)& $\mathcal{S} \circ \mathcal{F} \circ \mathcal{W}$\\
			\bottomrule
			\bottomrule
		\end{tabular}
	\end{table} 

By capitalizing on these strong commonalities, the open-source MATLAB\textsuperscript{\textregistered} library \globalbioim simplifies, unifies, and standardizes the resolution of inverse problems given by~\eqref{eq:OptiPb}. Hence, the \globalbioim toolbox gives access to state-of-the-art reconstruction algorithms usable in a wide range of imaging applications.
Its design is modular, with three main types of entities:
forward models, cost functions, and solvers. This permits the user to modify each component independently, which is crucial for the handling of a variety of imaging models and solvers within a common framework. This modularity also makes \globalbioim easily extensible to new modalities and novel reconstruction methods.

\globalbioim is distributed as an open-source MATLAB\textsuperscript{\textregistered} software. We expect it to respond to the needs of imaging scientists looking for reliable and easy-to-use computational tools for the reconstruction of their images. We also believe that \globalbioim will be of interest to developers of algorithms who focus on the mathematical and algorithmic details of the reconstruction methods. 

The present paper provides a functional description of the structure and the key components of the \globalbioim library. It completes and extends our previous brief communication~\cite{Unser2017}. For a detailed technical documentation, we refer the reader to an online documentation (\url{http://bigwww.epfl.ch/algorithms/globalbioim/}).

\subsection{Related Work}

The development of open-source libraries/toolboxes in imaging sciences has received considerable attention during the past two decades. The majority of existing softwares for solving inverse problems are dedicated to specific modalities, with various degrees of sophistication. Moreover, they cover the whole panel of programming languages.

There exists a large number of toolboxes dedicated to tomographic reconstruction for x-ray computed tomography, positron-emission tomography, single-photon-emission computed tomography,  or (scanning) transmission electron microscopy. These include among others ASTRA~\cite{astra}, CASToR~\cite{castor}, CONRAD~\cite{conrad}, RTK~\cite{rtk}, STIR~\cite{stir}, or TIGRE~\cite{tigre}.

For fluorescence microscopy, DeconvolutionLab \cite{sage2017deconvolutionlab2} provides a set of deconvolution methods that range from naive inverse filters to more sophisticated iterative approaches. The emergence of superresolution fluorescence microscopy techniques has also promoted the development of toolboxes tailored for their specific inverse problems. For instance, FairSIM~\cite{Mueller16} and Simtoolbox~\cite{Krizeka15} are dedicated to the reconstruction of structured-illumination microscopy data. For single-molecule localization microscopy, one can find dedicated localization plugins such as SMAP~\cite{li2018real} and ThunderSTORM~\cite{ovesny2014thunderstorm}.
  
Although dedicated to specific physical models, the aforementioned toolboxes generally rely on similar reconstruction methods, ranging from Wiener filtering to advanced regularized iterative algorithms. 
Conversely, libraries that are generic have recently also been designed to handle multiple imaging modalities. The LazyAlgebra toolbox \cite{eric_thiebaut_2018_1745422} provides an operator-algebra mechanism in Julia that can be combined with optimization packages for solving inverse problems. More complete libraries such as  AIR Tools  (MATLAB\textsuperscript{\textregistered}) \cite{hansen2018air}, IR Tools (MATLAB\textsuperscript{\textregistered}) \cite{gazzola2017ir}, or TiPi (Java\textsuperscript{\texttrademark}) \cite{eric_thiebaut_2018_1745424} provide elements for the implementation of forward models, as well as iterative solvers to tackle the associated inverse problems. The disadvantage of these toolboxes is that the optimization algorithms they provide are generally implemented to minimize a specific functional, thus limiting their modularity. 
 
In contrast, \globalbioim provides a fully modular environment where one can not only easily combine functionals and operators to define the loss to be minimized in~\eqref{eq:OptiPb}, but also benefit from a variety of solvers.  This philosophy is shared by a few other toolboxes with different programming languages, such as the Operator Discretization Library (in Python)~\cite{adler2017odl} and the Rice Vector Library (in C++)~\cite{padula2009software}.


\section{General Philosophy and Organization}

When tasked with the design of a reconstruction algorithm for a new imaging problem, the common practice follows a three steps process.
\begin{enumerate}
    \item  Modelization of the acquisition system $\Rightarrow$ Implementation of $\Hd$.
    \item  Formulation of the reconstruction as an optimization problem (\ie the cost function) $\Rightarrow$ Choice of $\mathcal{D}$ and $\mathcal{R}$ in~\eqref{eq:OptiPb}.
    \item  Deployment of an optimization method $\Rightarrow$ Choice of a solver for~\eqref{eq:OptiPb}.
\end{enumerate}

\begin{figure}
    \centering
    \begin{tikzpicture}
        \draw node [abstractClass] at (0,0) (map) {Map};
        \draw node [abstractClass] at (-3,-1.5) (linop) {LinOp};
        \draw node [abstractClass] at (3,-1.5) (cost) {Cost};
        \draw node [abstractClass] at (6.5,-1.5) (opti) {Opti};
        \draw[-latex,thick](cost.north)    -| (3,-0.75) -| (map);
        \draw[-latex,thick](linop.north)    -| (-3,-0.75) -| (map);
        \draw node [concreteClass] at (-3,-2.7) (linopconv) {LinOpConv};
        \draw node [concreteClass] at (-3,-3.9) (linopgrad) {LinOpGrad};
        \draw node [blank] at (-3,-5.1)  {\Large$\cdots$};
        \draw[latex-,thick] (linop.west) -| (-4.5,-1.5) -| (-4.5,-5.1);
        \draw[latex-,thick] (linop.west) -| (-4.5,-1.5) -| (-4.5,-3.9) -| (linopgrad.west);
        \draw[latex-,thick] (linop.west) -| (-4.5,-1.5) -| (-4.5,-2.7) -| (linopconv.west);
        \draw node [concreteClass] at (3,-2.7) (costl2) {CostL2};
        \draw node [concreteClass] at (3,-3.9) (costtv) {CostTV};
        \draw node [blank] at (3,-5.1) {\Large$\cdots$};
        \draw[latex-,thick] (cost.west) -| (1.5,-1.5) -| (1.5,-5.1);
        \draw[latex-,thick] (cost.west) -| (1.5,-1.5) -| (1.5,-3.9) -| (costtv.west);
        \draw[latex-,thick] (cost.west) -| (1.5,-1.5) -| (1.5,-2.7) -| (costl2.west);
        \draw node [concreteClass] at (6.5,-2.7) (optiFBS) {OptiFBS};
        \draw node [concreteClass] at (6.5,-3.9) (optiadmm) {OptiADMM};
        \draw node [blank] at (6.5,-5.1) {\Large$\cdots$};
        \draw[latex-,thick] (opti.west) -| (5,-1.5) -| (5,-5.1);
        \draw[latex-,thick] (opti.west) -| (5,-1.5) -| (5,-3.9) -| (optiadmm.west);
        \draw[latex-,thick] (opti.west) -| (5,-1.5) -| (5,-2.7) -| (optiFBS.west);
        \draw node [concreteClass] at (0,-2.7) (nonlinop) {OpEWSqrt};
        \draw[-latex,thick] (nonlinop.west) -| (-1.5,-2.7) -| (-1.5,-0.75) (map.south)  ;
         \draw node [blank] at (0,-3.9) {\Large$\cdots$};
        \draw[thick]  (-1.5,-3.9) -- (-1.5,-0.75) ;
    \end{tikzpicture}
    \caption{Hierarchy of classes in \globalbioim. Abstract classes are represented in blue and derived classes in gray. Nonlinear operators (such as the element-wise square-root \meth{OpEWSqrt})  directly inherit from the abstract \map  class.}
    \label{fig:Classes}
\end{figure}
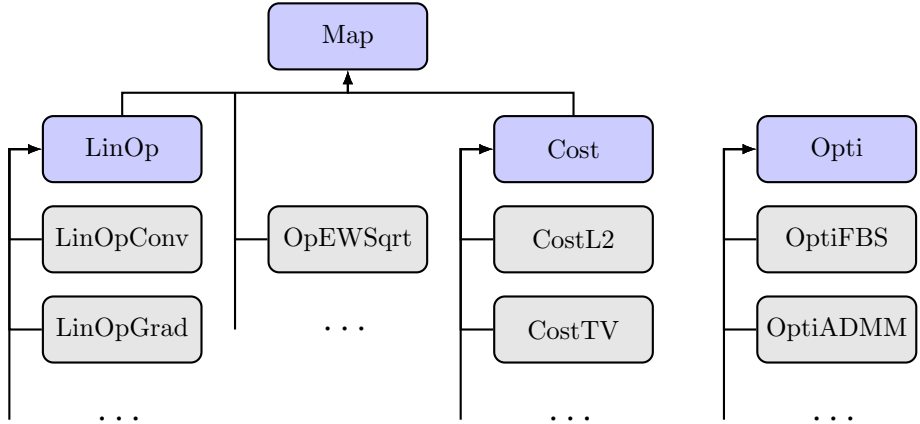

This standard pipeline motivates the organization of \globalbioim around three dedicated main abstract classes: \linop, \cost, and \opti. Because linear operators and cost functions both belong to the larger mathematical class of maps, the \linop and \cost classes are defined as particular instances of a generic abstract class \map. The latter also allows for a proper inclusion of nonlinear operators. The organization of the library is illustrated in Figure \ref{fig:Classes}. It is guided by five general principles.
\begin{itemize}
    \item \textbf{Modularity.} All objects are defined as individual modules that can be combined to generate a particular reconstruction workflow. Each building block can thus be easily changed to define new reconstruction pipelines.
    \item \textbf{Flexibility.} The constraints to fulfill during implementation are few. New objects can easily be plugged into the framework of \globalbioim. 
    \item \textbf{Abstraction.} The four abstract classes (\linop, \cost, \opti, \map) define a limited set of attributes and methods that are shared by their derived classes (\ie subclasses). This constitutes a common guideline for the implementation of subclasses. Moreover, generic concepts---basically, interactions between classes---are implemented at the level of the abstract classes and benefit directly to all subclasses.
    \item \textbf{Readability.} Reconstruction scripts are written in a way that mimics equations in scientific papers, hence keeping a simple connection between theory and implementation.
    \item \textbf{User-friendliness.}  The definition (or update) of a new subclass only requires one to create (or edit) a single file. Moreover, the usage of \globalbioim does not require one to understand advanced computing concepts. 
\end{itemize}


\section{Abstract Classes}

We now present the four abstract classes that build up the skeleton of the \globalbioim library. The methods within these abstract classes are prototypes that have to be implemented in derived classes. There are exceptions for some generic concepts (\eg the chain rule) that are directly implemented in the abstract classes. For the sake of conciseness,  we only review here the key attributes and methods of those classes. An exhaustive list of those features can be found in the online documentation (\url{http://bigwww.epfl.ch/algorithms/globalbioim/}) within the sections ``List of Methods'' and ``List of Properties''.

\subsection{\map Class} \label{sec:Map}

The  abstract \map class defines the basic attributes and methods of an operator $\Hd : \R^N \rightarrow \R^M$. These include, at the very minimum, the input size $N$, the output size $M$, and the method \meth{apply} that computes $\gd = \Hd\{\fd\}$ for a given $\fd \in \R^N$. 

In addition, because optimization algorithms may require the differentiation of the objective function in \eqref{eq:OptiPb}, the \map class defines the method \meth{applyJacobianT}. Given $\vd \in \R^M$ and $\fd \in \R^N$, this method computes $\ud = [\Jd_{\Hd}\{\fd\}]^T\vd$, where $\Jd_{\Hd}\{\fd\} \in \R^{M \times N}$ is the Jacobian matrix of $\Hd$ (assuming that the latter is differentiable). It is formed out of the first-order partial derivatives of the operator~$\Hd$, with
\begin{equation}
    [\Jd_{\Hd}\{\fd\}]_{m,n} =  \frac{\partial \Hd_m}{\partial \fd_n},
\end{equation}
where $\Hd_m : \R^N \rightarrow \R$ is such that $\Hd = [\Hd_1, \ldots, \Hd_M]^T$. Similarly, for invertible maps, the method \meth{applyInverse} allows one to compute $\fd = \Hd^{-1}\{\gd\}$ for $\gd \in \R^M$. 

In addition to the ``apply''-type methods, the \map class provides prototype methods prefixed by ``make''. These can be implemented in derived classes to create new instances of \map objects that are related to $\Hd$. For instance, the method \meth{makeInversion} returns a \map object that corresponds to $\Hd^{-1}$. 

The prototype methods are also used to overload the MATLAB\textsuperscript{\textregistered} operators ``$*$'' (\meth{mtimes}), ``$+$'' (\meth{plus}), and ``$-$'' (\meth{minus}). Hence, the composition  between two \map  objects can be specified easily as \meth{H = H1 * H2}. This will execute the method \meth{makeComposition} of \meth{H1} with \meth{H2} as its argument. By default,  the resulting \meth{H} will be a \meth{MapComposition} object that benefits from the generic implementations (\eg successive calls for \meth{apply}, chain rule for \meth{applyJacobianT}) provided in the  \meth{MapComposition} class. 

Similarly, the operators  ``$+$'' and ``$-$'' are associated to the \meth{MapSummation} class. It is noteworthy to mention that this default behavior can be specialized in derived classes with a proper implementation of the ``make'' methods. This results in automatic simplifications, as described in Section \ref{sec:AutomaticCompo}.

\subsection{\linop Class} \label{sec:LinOp}

A particular class of map objects contains linear operators $\Hd : \R^N \rightarrow \R^M$ that satisfy
\begin{equation}
    \Hd \{\alpha \fd +  \gd \} = \alpha \Hd \{ \fd \} +    \Hd \{ \gd \}
\end{equation}
for all scalar $\alpha$ and vectors $\fd \in \R^N$ and $\gd \in \R^N$. Linear operators are generally represented as a matrix $\Hd \in \R^{M\times N}$. They are widely used in practice to model imaging systems. In addition to being good approximators, they lead to convex optimization problems for which there exist efficient solvers. An important subclass is formed by the convolution operators that are implemented very efficiently using FFTs. All this motivates the definition of the \linop class, which inherits from the attributes and the methods of \map while defining novel ones. 

For linear operators, the transposed Jacobian matrix $[\Jd_{\Hd}\{\fd\}]^T$ is independent of $\fd$ and is equal to the adjoint operator $\Hd^T$. Thus, the \linop class provides  the method \meth{applyAdjoint} that computes $\ud = \Hd^T \vd$ for $\vd \in \R^M$ and is directly used  to implement the method \meth{applyJacobianT}.
Hence, to allow a \linop to be differentiated only requires implementation of \meth{applyAdjoint},
rather than \meth{applyJacobianT}.
In keeping with the aforementioned philosophy, the companion method  \meth{makeAdjoint} allows one to instantiate a new \linop corresponding to the adjoint $\Hd^T$. 

For least-squares minimization, the normal operators $\Hd^T \Hd$ and $\Hd \Hd^T$ are  at the core of many optimization algorithms. Hence, the methods \meth{applyHtH} and \meth{applyHHt} as well as their ``make'' counterparts are defined in the \linop class. They can be implemented in derived classes to provide implementations that are faster than the default successive application of $\Hd$ and $\Hd^T$. This is particularly useful when $\Hd^T \Hd$ turns out to be a convolution,
as is the case for deconvolution, cryo-electron microscopy~\cite{Donati2018,Vonesch2011}, and x-ray computed tomography~\cite{McCann2016}.


\subsection{\cost Class} \label{sec:Cost}

Cost functions are mappings for which $M=1$ (\ie $\mathcal{J} : \R^N \rightarrow \R$). Hence, the  abstract \cost class inherits from all the attributes and methods defined by the \map class. However, the \cost  class also defines new  attributes and methods that are specific to cost functions. For instance, the method \meth{applyProx} is dedicated to the computation of the proximity operator of $\mathcal{J}$, which is required for a broad range of optimization algorithms.
It is  defined by~\cite{Moreau1962} as
\begin{equation}\label{eq:prox}
    \mathrm{prox}_{\mathcal{J}}(\zd) = \argmin{\fd \in \R^N}{\left( \frac12 \|\fd- \zd\|_2^2 + \mathcal{J}(\fd)\right)}.
\end{equation}

The method \meth{applyGrad} computes the gradient $\nab \mathcal{J}$ of the functional $\mathcal{J}$. Similarly to the method \meth{applyAdjoint} for \linop, \meth{applyGrad}  can be seen as an alias for the method \meth{applyJacobianT}. This ensures consistency with the standard terminology employed in scientific publications.

\subsection{\opti Class} \label{sec:Opti}

The last abstract class \opti is a prototype for  optimization algorithms. Given a cost function $\mathcal{J}$ resulting from the composition/addition of \map, \linop, and \cost objects, the \meth{run} method of the \opti class implements a general iterative scheme to minimize $\mathcal{J}$. It includes calls to the methods \meth{initialize}, \meth{doIteration}, and \meth{updateParams}, which are implemented in every derived class. 

The method \meth{initialize} performs the computations required prior to starting the main loop of the iterative scheme. Then, \meth{doIteration} is executed at each iteration, preceded by a call to \meth{updateParams} that modifies the parameters of the algorithm (\eg by modifying the step size in a descent method). 

The convergence of the algorithm is monitored during the optimization using a \meth{TestCvg} object (set as an attribute of the \opti object). \globalbioim contains various \meth{TestCvg} classes that implement different convergence criteria (\eg \meth{TestCvgStepRelative}, \meth{TestCvgCostRelative}). They can also  be combined using the class \meth{TestCvgCombine}. Finally, the verbose output is controlled by an \meth{OutputOpti} object (again, set as an attribute of the \opti object). Hence, one can easily tune the information displayed and saved during iteration by defining a custom \meth{OutputOpti} class. 
  
  
\section{Key Features of the Library} 

In this section, we highlight some of the most remarkable features of \globalbioim, which are intended to simplify the development process.

\subsection{Interface and Core Methods}

  \map, \linop, and \cost classes contain two types of methods, which come in pairs. \textit{Interface methods} are only implemented in abstract classes and cannot be overridden in derived classes (sealed methods). However, they can be executed by an instantiated object of the class. On the other hand, \textit{core methods} are not implemented in abstract classes, but in derived classes only. In addition, they cannot be executed by an instantiated object (private methods). 
  
  This scheme allows for the separation of preprocessing computations, which are common to all derived classes, from the core computations of the method, which are class-dependent. When executed, an interface method checks that inputs are the correct size prior to executing the associated core method. Interface methods are also used to manage the memoize mechanism (see Section \ref{sec:MemVsCost}).
  
  From a user viewpoint, only core methods matter. They have to be implemented in derived classes without having to deal with input checking and the memoize mechanism. In the library, the core methods differ from the interface methods by the suffix ``\_'' (\eg \meth{apply\_} versus  \meth{apply}).

\subsection{Composition of Operators and Automatic Simplification} \label{sec:AutomaticCompo}

Up to now, the reader may wonder why it is useful to allow ``make''-type methods to be overloaded in derived classes. Actually, these methods are the key ingredients for the automatic simplification mechanism deployed by \globalbioim. When compositions between maps occur,
they allow for the instantiation of specific classes instead of the default generic classes such as \meth{MapComposition}, \meth{MapInversion}, or \meth{MapSummation}. Consequently, the resulting object generally enjoys faster implementations.

To illustrate this feature, let us consider a convolution operator $\Hd$. Its adjoint $\Hd^T$ and the normal operator $\Hd^T \Hd$ turn out to be convolution operators as well, whose kernels can be precomputed from that of $\Hd$. Hence, the \meth{LinOpConv} class implements the \meth{makeAdjoint\_} and \meth{makeHtH\_} methods by instantiating a new \meth{LinOpConv} with the adequate kernel. 
\begin{lstlisting}
function M = makeAdjoint_(this)
    % Reimplemented from parent class :class:`LinOp`.
    M=LinOpConv(conj(this.mtf),this.isReal,this.index);
end
function M = makeHtH_(this)
    % Reimplemented from parent class :class:`LinOp`.
    M=LinOpConv(abs(this.mtf).^2,this.index);
end
\end{lstlisting}
Because \meth{makeAdjoint\_} and \meth{makeHtH\_} are used to overload the operators ``$'$'' and ``$*$'', respectively, we obtain the following automatic simplification.
\begin{lstlisting}
>> H=LinOpConv(fftn(psf));
>> L=H'*H
L =
  LinOpConv with attributes:
      mtf: [256x256 double]
      ...
\end{lstlisting}

There are many more examples of ``make'' methods in  \globalbioim  (see also the methods \meth{plus\_} and \meth{mpower\_}). For instance, the  \meth{LinOpConv} class provides the following implementation for the method \meth{plus\_} (which is used to overload ``$+$'').
\begin{lstlisting}[linebackgroundcolor={%
                                \color{lightgrey}
                                \ifnum\value{lstnumber}=3 \color{orange!30} \fi
                                \ifnum\value{lstnumber}=4 \color{orange!30} \fi
                                }]
function M = plus_(this,G)
   % Reimplemented from parent class :class:`LinOp`.
   if isa(G,'LinOpDiag') && G.isScaledIdentity     % If sum with a constant diagonal operator
      M=LinOpConv(G.diag+this.mtf,this.isReal,this.index);
   elseif isa(G,'LinOpConv')                       % If sum with a convolution operator
      M=LinOpConv(this.mtf+G.mtf,this.isReal,this.index);
   else                                            % Otherwise call superclass plus_ method
      M=plus_@LinOp(this,G);
   end
end
\end{lstlisting}
This allows the following simplification.
\begin{lstlisting}
>> H=LinOpConv(fftn(psf));
>> I=LinOpIdentity(H.sizein);
>> L=H'*H + I
L =
  LinOpConv with attributes:
      mtf: [256x256 double]
      ...
\end{lstlisting}

We conclude this section with two examples that demonstrate the relevance of this automatic simplification mechanism.

\begin{example}[Proximity operator with semiorthogonal linear transform] \label{ex:semiOrtho}
 Let $\mathcal{J}$ be a lower semicontinuous convex functional and $\Ld$ be a semiorthogonal linear operator (\ie $\Ld \Ld^T = \nu \Id$ for $\nu >0$). Then, as demonstrated in \cite[Lemma 2.4]{combettes2008proximal}, the proximity operator of $\mathcal{J}(\Ld \cdot)$ is given by
\begin{equation}\label{eq:ProxOrtho}
    \mathrm{prox}_{\alpha \mathcal{J}(\Ld \cdot)}(\zd) = \zd + \nu^{-1} \Ld^T \left( \mathrm{prox}_{\nu \alpha \mathcal{J}}(\Ld \zd) - \Ld \zd \right).
\end{equation}
Due to the automatic simplification mechanism of \globalbioim, one can easily verify whether $\Ld$ is semiorthogonal in the constructor of the \meth{CostComposition} class ($\Ld \rightarrow$ \meth{this.H2}).
{\normalfont
\begin{lstlisting}
T=this.H2*this.H2';                                           % Build L'*L
if isa(T,'LinOpDiag') && T.isScaledIdentity && T.diag>0       % Check if scaled identity
    this.isH2SemiOrtho=true;
    this.nu=T.diag;
end
\end{lstlisting}}
\noindent Then, \eqref{eq:ProxOrtho} is exploited in the implementation of the \meth{applyProx\_} method of the \meth{CostComposition} class ($\mathcal{J} \rightarrow$ \meth{this.H1}).
{\normalfont
\begin{lstlisting}[linebackgroundcolor={%
                                \color{lightgrey}
                                \ifnum\value{lstnumber}=3 \color{orange!30} \fi
                                \ifnum\value{lstnumber}=4 \color{orange!30} \fi
                                }]
function x=applyProx_(this,z,alpha)
    if this.isConvex && this.isH2LinOp && this.isH2SemiOrtho
        H2z=this.H2*z;
        x = z + 1/this.nu*this.H2.applyAdjoint(this.H1.applyProx(H2z,alpha*this.nu)-H2z);
    else
        x = applyProx_@Cost(this,z,alpha);
    end
end
\end{lstlisting}
}
\noindent As a result, the composition of a \cost object that has an implementation of its proximity operator with a semi-orthogonal \linop object automatically leads to a \meth{CostComposition} object that has an implementation of the \meth{applyProx\_} method. 
\end{example}

\begin{example}[Woodbury matrix identity]
Let $\mathcal{J}(\fd) = \frac12 \| \Sd \Hd \fd - \gd \|_2^2$, where $\Sd$ is a downsampling operator and $\Hd$ is a convolution operator. It follows from \eqref{eq:prox} that
\begin{align}
    \mathrm{prox}_{\alpha \mathcal{J}}(\ud) & = \left( \alpha \Hd^T \Sd^T \Sd \Hd + \Id \right)^{-1}\left(\alpha \Hd^T \Sd^T \gd + \ud \right) \nonumber  \\
    & = \left(\Id - \alpha \Hd^T \Sd^T \left( \Id + \alpha \Sd \Hd \Hd^T \Sd^T  \right)^{-1} \Sd \Hd \right) \left(\alpha \Hd^T \Sd^T \gd + \ud \right), \label{eq:Woodbury2}
\end{align}
where the Woodbury matrix identity \cite{Hager89} is used to get \eqref{eq:Woodbury2}. Moreover, it turns out that $\alpha \Sd \Hd \Hd^T \Sd^T$ is a convolution operator \cite[Lemma A.3]{Soubies18} and, thus, that $\left( \Id + \alpha \Sd \Hd \Hd^T \Sd^T  \right)$ can easily be inverted in the Fourier domain. In order to apply~\eqref{eq:Woodbury2}, the specification of $\alpha \Sd \Hd \Hd^T \Sd^T$ as a convolution is implemented in \globalbioim. This is done in the \meth{makeComposition\_} method of \meth{LinOpDownsample}, which returns a \meth{LinOpConv} object when appropriate.
{\normalfont
\begin{lstlisting}[linebackgroundcolor={%
                                \color{lightgrey}
                                \ifnum\value{lstnumber}=8 \color{orange!30} \fi
                                }]
function G = makeComposition_(this, H)
    % Reimplemented from parent class :class:`LinOp`
    % ...
    if isa(H, 'LinOpComposition')
        if isa(H.H2,'LinOpAdjoint') && isequal(H.H2.TLinOp,this)
            if isa(H.H1, 'LinOpConv')
                P=LinOpSumPatches(this.sizein,this.sizein./this.df);
                G = LinOpConv(P*H.H1.mtf/prod(this.df),H.H1.isReal); 
    % ...
\end{lstlisting}}
\noindent As a result, $\left( \Id + \alpha \Sd \Hd \Hd^T \Sd^T  \right)$ is identified as being an invertible operator, and \eqref{eq:Woodbury2} can be directly implemented as follows. 
{\normalfont
\begin{lstlisting}[linebackgroundcolor={%
                                \color{lightgrey}
                                \ifnum\value{lstnumber}=3 \color{orange!30} \fi
                                }]
H=LinOpConv(fftn(psf)); S=LinOpDownsample(H.sizein,[2,2]); fwd=S*H;     % Forward model
In=LinOpIdentity(S.sizein); Im=LinOpIdentity(S.sizeout);                   % Identity operators
prox=(In - alpha*fwd'*(Im + alpha*(fwd*fwd'))^(-1)*fwd)*(alpha*fwd'*g+u);
\end{lstlisting}}
\noindent A complete script (\text{TestProxL2DownSampledConv}) in which this example is implemented can be found in the folder \text{Cost/Tests/} of the \globalbioim library.
\end{example}

\subsection{The Memory versus Computational Cost Dilemma}\label{sec:MemVsCost}

  The computation and storage of fixed quantities can significantly accelerate iterative reconstruction methods. For instance, let us consider the least-squares functional $\mathcal{J}(\fd)= \frac12 \| \Hd \fd - \gd \|^2_2$, where $\Hd$ is a convolution operator. Then, the minimization of $\mathcal{J}$ by a gradient-descent algorithm requires the evaluation of
  \begin{align}
      \nab \mathcal{J}(\fd) & = \Hd^T (\Hd \fd - \gd) \label{eq:gradLS_1}\\
      & =  \Hd^T\Hd \fd - \Hd^T\gd \label{eq:gradLS_2}
  \end{align}
  at each iteration. The computational burden of this operation is directly related to the implementation strategy. The formulation in~\eqref{eq:gradLS_1} requires the evaluation of both $\Hd$ and $\Hd^T$, leading to the overall cost of two FFTs plus two iFFTs. Instead, since $\Hd^T \Hd$ turns out to be a convolution in this example, the formulation in \eqref{eq:gradLS_2} opens the door to a faster computation of $\nab \mathcal{J}$. The price to pay, however, is storage for the quantity $\Hd^T\gd$. Imposing one of the above implementations to users could lead to severe memory issues or extremely slow computations, depending on the considered problem and the available hardware resources. Therefore, in \globalbioim, the choice between speed and memory consumption is left to the user by means of the Boolean attribute \meth{doPrecomputation} of the abstract class \map. When activated, the instanciated object is allowed to store relevant quantities for acceleration purposes, at the expense of larger memory consumption. For instance, consider the \cost object corresponding to $\mathcal{J}= \frac12 \| \Hd \cdot - \gd \|^2_2$ for which the \meth{doPrecomputation} option is activated.
  \begin{lstlisting}[linebackgroundcolor={%
                                \color{lightgrey}
                                \ifnum\value{lstnumber}=4 \color{orange!30} \fi
                                }]
>> H=LinOpConv(fftn(psf));        % Convolution operator with a 256x256x256 kernel
>> L2=CostL2([],y);               % L2 cost function
>> J=L2*H;                        % Composition of L2 with H 
>> J.doPrecomputation=true;       % Activation of the doPrecomputation option
  \end{lstlisting}
  Then, we evaluate the gradient of $\mathcal{J}$ at the two random points $\fd_1$ and $\fd_2$.
  \begin{lstlisting}
>> f1=rand(J.sizein);f2=rand(J.sizein);      % Generation of the random points f1 and f2
>> tic; g=J.applyGrad(f1); toc;              % Computation of the gradient at f1
Elapsed time is 1.870925 seconds.
>> tic; g=J.applyGrad(f2); toc;              % Computation of the gradient at f2
Elapsed time is 0.942075 seconds.
  \end{lstlisting}
  We observe that the second gradient computation is twice as fast as the first one. This is because the quantity $\Hd^T \gd$ is computed and stored at the first call of the \meth{applyGrad} method. For all subsequent calls, the computational burden is reduced to the application of $\Hd^T \Hd$ in \eqref{eq:gradLS_2}.

  Another feature that allows for faster computations at the expense of larger memory consumption is provided by the structure attribute \meth{memoizeOpts} of the abstract class \map. 
  When this attribute is activated, both the input and the result of the evaluation are stored whenever the corresponding \map object is evaluated. Hence, if the object is subsequently evaluated with the same input, the stored result is returned without any computation.
  
  \begin{lstlisting}[linebackgroundcolor={%
                                \color{lightgrey}
                                \ifnum\value{lstnumber}=2 \color{orange!30} \fi
                                }]
>> H=LinOpConv(fftn(psf));               % Convolution operator with a 256x256x256 kernel
>> H.memoizeOpts.apply=true;             % Activation of memoize for the apply method 
>> f1=rand(H.sizein);f2=rand(H.sizein);  % Generation of the random points f1 and f2
>> tic; g=H*f1; toc;                     % First H*f1
Elapsed time is 0.822508 seconds.
>> tic; g=H*f1; toc;                     % Second (consecutive) H*f1. Returns the stored result
Elapsed time is 0.020624 seconds.
>> tic; g=H*f2; toc;                     % Computation for the new input f2
Elapsed time is 0.823498 seconds.
  \end{lstlisting}
  This option proves to be particularly useful to avoid multiple computations within iterative methods. For instance, at each iteration of \meth{OptiVMLMB}, both the cost $\mathcal{J}$ and its gradient $\nab \mathcal{J}$ need to be evaluated at the same point $\fd$. For least-squares minimization, this involves computing $\mathcal{J}(\fd) = \frac12 \| \Hd \fd - \gd \|_2^2$ and $\nab \mathcal{J}(\fd) = \Hd^T (\Hd \fd - \gd)$, which both call for the quantity $\Hd \fd$. Hence, activating the \meth{memoize} option for the \meth{apply} method of $\Hd$ allows for the savings of one evaluation of $\Hd \fd$ per iteration.

\subsection{GPU Computing} \globalbioim provides two functions that allow the user to easily run any 
 reconstruction pipeline on the GPU for faster computation. The function \meth{useGPU}, which is typically called at the beginning of the script, allows selection of the computation mode: CPU computation (default), GPU computation with the MATLAB\textsuperscript{\textregistered} Parallel Computing Toolbox\textsuperscript{\texttrademark}, or GPU computation with CudaMat (\url{https://github.com/RainerHeintzmann/CudaMat}). Next, the function \meth{gpuCpuConverter} converts the input variable to the appropriate data type as specified by \meth{useGPU}. A typical use of these functions is presented below.
 \begin{lstlisting}
useGPU(1);               % Set the GPU mode to 1 -> Matlab Parallel Computing Toolbox (TM)

%-- Load data
load('data');            % Variable g
g=gpuCpuConverter(g);    % Convert them to the correct type
% ... load and convert other variables
 \end{lstlisting}


\section{An Example with MultiChannel Deconvolution} \label{sec:appli}

\subsection{Simulation Setting} 

\begin{figure}
    \centering
	\input{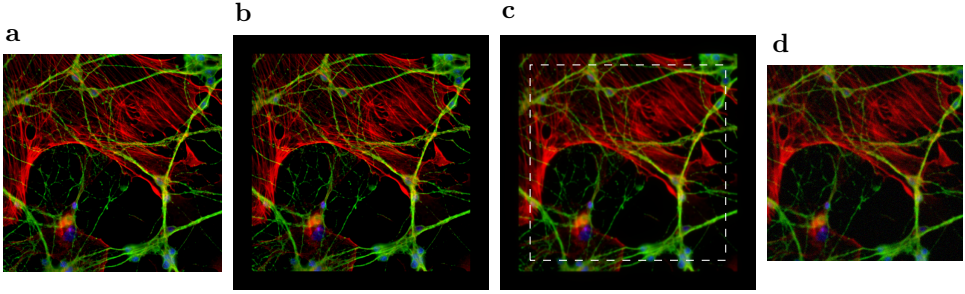}
    \caption{Simulation of multichannel blurred data. a)~Input image ($512 \times 512 \times 3$). b)~The image is zero-padded. c)~Each channel is convolved with its corresponding PSF and a central region of size $(460 \times 460 \times 3)$ is extracted (simulated field-of-view). d)~Data are corrupted by additive Gaussian noise so that the resulting signal-to-noise ratio (SNR) is equal to 10dB.}
    \label{fig:simuData}
\end{figure}

We consider the sample  depicted in Figure~\ref{fig:simuData}a. It has been extracted from the neuronal culture acquisition shared by Schmoranzer  on the Cell Image Library website (\url{http://cellimagelibrary.org/images/41649}). It contains three channels that we process independently. Our simulation pipeline is illustrated in Figure~\ref{fig:simuData}. It encompasses several steps to account for the fact that, for real-world experiments, (i) the underlying  sample is generally not supported within the field-of-view of the microscope; (ii) the sample is not periodic (contrary to what is implicitly assumed when using FFTs to perform the convolution). The three point-spread functions (PSF) have been generated in the Fourier domain. They are related to the classical Airy disk model, which is a radial function with the profile 
\begin{equation}
     h(\rho) = \left\lbrace  
     \begin{array}{ll}
       \disp \frac{1}{\pi} \left(2 \cos^{-1}\left(\frac{\rho}{\rho_c}\right) - \sin \left(2 \cos^{-1}\left(\frac{\rho}{\rho_c}\right)\right) \right),  &  \forall \rho < \rho_c\\
       0,   & \text{ otherwise},
     \end{array}
     \right.
\end{equation}
where $\rho_c = 2 \mathrm{NA}/\lambda_{\mathrm{exc}}$ is the cutoff frequency which depends on the numerical aperture $\mathrm{NA}$ and the excitation wavelength $\lambda_{\mathrm{exc}}$. Here, we set $\mathrm{NA}=1.4$ and $\lambda_{\mathrm{exc}}$ to 654\nano\meter~(CY3 dye, red), 542\nano\meter~(FITC dye, green), and 477\nano\meter~ (DAPI dye, blue) for the three channels, respectively. Finally, the spatial sampling step (\ie the camera pixel size) is set to 64.5\nano\meter. Note that the data generated with this pipeline contain contributions from structures that lie outside the field-of-view.

\subsection{Deconvolution} \label{sec:deconv}

Given the blurred and noisy data $\{\gd_k\in \R^M\}_{k=1}^3$, we formulate the deconvolution task as the  optimization problem
\begin{equation}\label{eq:OptiPbDeconv}
    \{\hat{\fd}_k\}_{k=1}^3 = \argmin{\{\fd_k \in \R^{N}\}_{k=1}^3}{\left( \sum_{k=1}^3 \frac12 \| \Sd \Hd_k \fd_k - \gd_k\|^2_2 + \lambda \mathcal{R}(\Ld \fd_k) + i_{\geq 0}(\fd_k) \right)},
\end{equation}
where $\Hd_k \in \R^{N \times N}$, $k \in \{1, 2, 3\}$, is the convolution operator for the $k$th channel, and $\Sd \in \R^{M\times N}$ selects the region of $\Hd \fd$ that corresponds to the field-of-view. Indeed, since the sample is not fully included in the field-of-view, we seek a wider reconstruction that is larger than the field-of-view (\ie $N>M$) in order to avoid reconstruction artifacts~\cite{Almeida2013,sage2017deconvolutionlab2}. Finally, $i_{\geq 0}(\fd) = \{ 0 \text{ if } \fd \in \R_{\geq 0}^N; + \infty \text{ otherwise}\}$  is a nonnegativity constraint.

With \globalbioim, the construction of the operators $\Hd_k$, $k \in \{1, \ldots, 3\}$, and $\Sd$ is done as follows.
\begin{lstlisting}
%-- Load Data --
load('psf'); szin=size(psf);   % Variable psf (512x512x3)
load('ground_truth');          % Variable gt (460x460x3)
load('data'); szout=size(g);   % Variable g (460x460x3) 

%-- Forward Model --
H=LinOpConv(fft2(psf),1,[1 2]); 
S=LinOpSelectorPatch(szin,[1 1 1],szout); 
\end{lstlisting}   
Here, the three PSFs are stacked within the same \meth{LinOpConv} operator which is set by the argument \meth{[1 2]} to apply a convolution only to the first two dimensions. Hence, it performs independent 2D convolutions for each channel. The selector operator $\Sd$ then extracts a region that has the same size as the data.

Next, both  the data fidelity term (\ie $\frac12 \| \cdot - \gd \|_2^2$) and the nonnegativity constraint (\ie $i_{\geq 0}$) can be defined with two lines of code.
\begin{lstlisting}
%-- L2 Loss function and nonnegativity constraint
L2=CostL2([ ],g);
P=CostNonNeg(szin);
\end{lstlisting}  

For the regularization term $\mathcal{R}(\Ld \cdot)$, we propose to illustrate the modularity of \globalbioim by providing a set of examples (see Figures \ref{fig:ListingNonDiff} and \ref{fig:ListingDiff}) that include various  regularizers.
\begin{itemize}
    \item The total-variation (TV) \cite{chambolle2010introduction,chambolle1997image,Rudin1992} combines the gradient operator $\Ld=  [\Dd_1 \, \Dd_2]^T$ with the ($\ell_{2},\ell_1$)-mixed norm  $\mathcal{R}=\| \cdot \|_{2,1}$. More precisely, for $\fd \in \R^N$, we have that
    \begin{equation}
       \mathcal{R}(\Ld \fd) = \sum_{n=1}^N \sqrt{[\Dd_1 \fd]_n^2 + [\Dd_2 \fd]_n^2},
    \end{equation}
    where $\Dd_1$ ($\Dd_2$, respectively) is the finite-difference operator along the first (second, respectively) dimension.
    \item The Hessian-Schatten-norm (HS) \cite{Lefkimmiatis13b,Lefkimmiatis2013}  computes the ($*,\ell_1$)-mixed norm $\mathcal{R} = \| \cdot \|_{*,1}$ of the Hessian operator $\Ld =  [\Dd_{ij}]_{1\leq i,j \leq 2}$ applied to $\fd \in \R^N$ as
     \begin{equation}
       \mathcal{R}(\Ld \fd) = \sum_{n=1}^N {\left\|  \begin{bmatrix} [\Dd_{11} \fd]_n & [\Dd_{12} \fd]_n \\ [\Dd_{21} \fd]_n & [\Dd_{22} \fd]_n \end{bmatrix} \right\|_{*}},
    \end{equation}
    where $\| \cdot \|_{*}$ denotes the nuclear norm (\ie the first-order Schatten norm). It  is defined as the $\ell_1$-norm of the singular values of its argument. Finally, $\Dd_{ij}$ denotes the operator of second-order finite difference  along the dimensions $i$ and $j$.
    \item The smoothed total-variation (S-TV) \cite{aujol2009some,chambolle2010introduction}  is defined, for $\varepsilon >0$, by
    \begin{equation}
       \mathcal{R}(\Ld \fd) = \sum_{n=1}^N \sqrt{[\Dd_1 \fd]_n^2 + [\Dd_2 \fd]_n^2 + \varepsilon^2}.
    \end{equation}
    \item The Good's roughness (GR) \cite{Verveer1999} is defined, for $\varepsilon >0$, by
     \begin{equation}
       \mathcal{R}(\Ld \fd) = \sum_{n=1}^N \frac{[\Dd_1 \fd]_n^2 + [\Dd_2 \fd]_n^2}{\sqrt{|\fd_n|^2 + \varepsilon^2}}.
    \end{equation}
\end{itemize}

Since the TV and HS regularizers are not differentiable, gradient-based methods cannot be used to minimize the objective function \eqref{eq:OptiPbDeconv}. However, for both these regularizers, the proximity operator of $\mathcal{R}(\cdot)$ can be efficiently computed (see \cite{combettes2008proximal}  for  $\| \cdot \|_{2,1}$ and \cite{chierchia2014nonlocal,Lefkimmiatis13b} for $\| \cdot \|_{\mathcal{S}_1,1}$). Hence, the optimization problem can be tackled using proximal-splitting algorithms such as the alternating direction method of multipliers (ADMM) \cite{afonso2011augmented,Boyd11,Fortin2000,Setzer2010} or the primal-dual method proposed in \cite{Condat13}. These algorithms are designed to minimize cost functions of the form $\mathcal{J}=\sum_{p=1}^P \mathcal{J}_p(\Td_p \cdot )$, where $\{\Td_p\}_{p=1}^P$ are linear operators and the $\{\mathcal{J}_p\}_{p=1}^P$ are ``simple" functions in the sense that their proximity operator can be evaluated efficiently. 

The scripts provided in Figure~\ref{fig:ListingNonDiff} illustrate how these two algorithms can be implemented within the framework of \globalbioim to solve Problem \eqref{eq:OptiPbDeconv} with TV or HS regularization. The modified lines of code between each setting have been highlighted---observe that very few modifications are needed. This underlines the simplicity of changing the regularizer and/or the algorithm within the \globalbioim framework.

For both algorithms, the splitting strategy is specified by the two cell arrays \meth{Fn} and \meth{Hn}. Note that, since $\Sd$ is a semi-orthogonal linear operator, the composition \meth{L2*S} results in a \cost object that has an implementation of the proximity operator (see Example \ref{ex:semiOrtho}). Moreover, the two scripts that use the primal-dual method illustrate the relevance of the automatic simplification features described in Section \ref{sec:AutomaticCompo}.
\begin{figure}[t]
    \centering
    \lstset{linewidth=0.47\textwidth}

\begin{center}
\hspace{-0.5cm}
    \begin{tikzpicture}
        \node (ADMM_TV) at (0,0) {
        \begin{lstlisting}
%-- Regularizer ---------------
lamb=5e-3;
L=LinOpGrad(szin,[1 2]);
R=CostMixNorm21(L.sizeout,4);

%-- Optimization Algorithm --------
Fn={L2*S,lamb*R,P};
Hn={H,L,LinOpIdentity(szin)};
rho_n=[1,1,1]*5e-1;                    
Opt=OptiADMM([],Fn,Hn,rho_n);          
Opt.OutOp=OutputOpti(1,S'*gt,50);      
Opt.CvOp=TestCvgStepRelative(1e-4);  
Opt.maxiter=500; Opt.ItUpOut=50; 
Opt.run(S'*y);
        \end{lstlisting}        
        };
        \node (ADMM_Hess) at (0.5\textwidth,0) {
        \begin{lstlisting}[linebackgroundcolor={%
                                \color{lightgrey} 
                                \ifnum\value{lstnumber}=3 \color{orange!30} \fi
                                \ifnum\value{lstnumber}=4 \color{orange!30} \fi}]
%-- Regularizer ---------------
lamb=5e-3;
L=LinOpHess(szin,[ ],[1 2]);
R=CostMixNormSchatt1(L.sizeout,1);

%-- Optimization Algorithm --------
Fn={L2*S,lamb*R,P};
Hn={H,L,LinOpIdentity(szin)};
rho_n=[1,1,1]*5e-1;                    
Opt=OptiADMM([],Fn,Hn,rho_n);       
Opt.OutOp=OutputOpti(1,S'*gt,50);      
Opt.CvOp=TestCvgStepRelative(1e-4);  
Opt.maxiter=500; Opt.ItUpOut=50; 
Opt.run(S'*y);
        \end{lstlisting}
        };
        \node (PD_TV) at (0.1,-5.2) {
        \begin{lstlisting}[linebackgroundcolor={%
                                \color{lightgrey} 
                                \ifnum\value{lstnumber}=9 \color{orange!30} \fi
                                \ifnum\value{lstnumber}=10 \color{orange!30} \fi
                                \ifnum\value{lstnumber}=11 \color{orange!30} \fi}]
%-- Regularizer ---------------
lamb=5e-3;
L=LinOpGrad(szin,[1 2]);
R=CostMixNorm21(L.sizeout,4);

%-- Optimization Algorithm --------
Fn={L2*S,lamb*R,P};
Hn={H,L,LinOpIdentity(szin)};
Opt=OptiPrimalDualCondat([ ],[ ],Fn,Hn);           
T=H'*H+L'*L+LinOpIdentity(szin); 
Opt.tau=1;Opt.sig=1/(Opt.tau*T.norm);      
Opt.OutOp=OutputOpti(1,S'*gt,50);      
Opt.CvOp=TestCvgStepRelative(1e-4);  
Opt.maxiter=500; Opt.ItUpOut=50; 
Opt.run(S'*y);
        \end{lstlisting}
        };
        \node (PD_Hess) at (0.507\textwidth,-5.2) {
        \begin{lstlisting}[linebackgroundcolor={%
                                \color{lightgrey} 
                                \ifnum\value{lstnumber}=3 \color{orange!30} \fi
                                \ifnum\value{lstnumber}=4 \color{orange!30} \fi
                                \ifnum\value{lstnumber}=9 \color{orange!30} \fi
                                \ifnum\value{lstnumber}=10 \color{orange!30} \fi
                                \ifnum\value{lstnumber}=11 \color{orange!30} \fi}]
%-- Regularizer ---------------
lamb=5e-3;
L=LinOpHess(szin,[ ],[1 2]);
R=CostMixNormSchatt1(L.sizeout,1);

%-- Optimization Algorithm --------
Fn={L2*S,lamb*R,P};
Hn={H,L,LinOpIdentity(szin)};
Opt=OptiPrimalDualCondat([ ],[ ],Fn,Hn);            
T=H'*H+L'*L+LinOpIdentity(szin); 
Opt.tau=1;Opt.sig=1/(Opt.tau*T.norm);            
Opt.OutOp=OutputOpti(1,S'*gt,50);      
Opt.CvOp=TestCvgStepRelative(1e-4);  
Opt.maxiter=500; Opt.ItUpOut=50; 
Opt.run(S'*y);
        \end{lstlisting}
        };
    \node[anchor=south] at (ADMM_TV.north) {Total-variation \cite{chambolle2010introduction,chambolle1997image,Rudin1992}};
    \node[rotate=90,anchor=south] at (ADMM_TV.west) {ADMM \cite{afonso2011augmented,Boyd11,Fortin2000,Setzer2010}};
    \node[anchor=south] at (ADMM_Hess.north) {Hessian-Schatten \cite{Lefkimmiatis13b,Lefkimmiatis2013}};
    \node[rotate=90,anchor=south] at (PD_TV.west) {Primal-dual \cite{Condat13}};
    \end{tikzpicture}
\end{center}
    \caption{\globalbioim scripts for minimizing \eqref{eq:OptiPbDeconv} with non-differentiable regularizers $\mathcal{R}(\Ld \cdot)$. Differences with respect to the script corresponding to ADMM with TV are highlighted.}
    \label{fig:ListingNonDiff}
\end{figure}
In order to ensure the convergence of the algorithm, the two parameters $\sigma>0$ and $\tau>0$ have to be chosen so that $\tau \sigma \|\sum_{p=1}^P \Td_p^T \Td_p \| \leq 1$ holds true \cite{Condat13}. The norm which is involved in this inequality is easily obtained with \globalbioim by building the operator \meth{T=H'*H + L'*L + LinOpIdentity(szin)} explicitly and computing its norm \meth{T.norm}. The key is that the composition used to build \meth{T} is automatically simplified to a convolution operator with the proper kernel.
Finally, observe that, as for the convolution operator, the gradient and Hessian operators are defined using the argument \meth{[1 2]}, which implies that these operators are applied independently to each channel.

As opposed to TV and HS, the S-TV and the GR regularizers are differentiable. Hence, the optimization problem in \eqref{eq:OptiPbDeconv} can be addressed through gradient-based methods. In Figure \ref{fig:ListingDiff}, we present scripts in which the objective function in \eqref{eq:OptiPbDeconv} with S-TV or GR regularization is minimized using either the variable-metric limited-memory-bounded (VMLMB) algorithm \cite{thiebaut2002optimization} or the fast iterative shrinkage-thresholding algorithm (FISTA) \cite{beck2009fast}.
For these two minimization algorithms, each iteration requires the evaluation of the gradient of $\sum_{k=1}^3 \| \Sd \Hd_k \cdot - \gd_k \|_2^2 + \lambda \mathcal{R}(\Ld \cdot)$ as well as a projection onto the set of nonnegative vectors. 
Once again, changing the regularizer or the optimization method only requires the modification of very few lines, as highlighted in Figure \ref{fig:ListingDiff}.

\begin{figure}
    \centering
\lstset{linewidth=0.47\textwidth}
\begin{center}
\hspace{-0.5cm}
    \begin{tikzpicture}
        \node (VMLMB_TVs) at (0,0) {
        \begin{lstlisting}
%-- Regularizer ---------------
lamb=5e-3;
L=LinOpGrad(szin,[1 2]);
RL=CostHyperBolic(L.sizeout,1e-7,4)*L;

%-- Optimization Algorithm --------
C = L2*S*H + lamb*RL;
Opt=OptiVMLMB(C,0.,[]);
Opt.OutOp=OutputOpti(1,S'*gt,50);      
Opt.CvOp=TestCvgStepRelative(1e-4);  
Opt.maxiter=500; Opt.ItUpOut=50; 
Opt.run(S'*y);
        \end{lstlisting}        
        };
        \node (VMLMB_GR) at (0.5\textwidth,0) {
        \begin{lstlisting}[linebackgroundcolor={%
                                \color{lightgrey}
                                \ifnum\value{lstnumber}=4 \color{orange!30} \fi
                                }]
%-- Regularizer ---------------
lamb=5e-3;
L=LinOpGrad(szin,[1 2]);
RL=CostGoodRoughness(L,1e-2);

%-- Optimization Algorithm --------
C = L2*S*H + lamb*RL;
Opt=OptiVMLMB(C,0.,[]);
Opt.OutOp=OutputOpti(1,S'*gt,50);      
Opt.CvOp=TestCvgStepRelative(1e-4);  
Opt.maxiter=500; Opt.ItUpOut=50; 
Opt.run(S'*y);
        \end{lstlisting}
        };
        \node (FISTA_TVs) at (0,-4.9) {
        \begin{lstlisting}[linebackgroundcolor={%
                                \color{lightgrey}
                                \ifnum\value{lstnumber}=8 \color{orange!30} \fi
                                \ifnum\value{lstnumber}=9 \color{orange!30} \fi
                                }]
%-- Regularizer ---------------
lamb=5e-3;
L=LinOpGrad(szin,[1 2]);
RL=CostHyperBolic(L.sizeout,1e-7,4)*L;

%-- Optimization Algorithm --------
C = L2*S*H + lamb*RL;
Opt=OptiFBS(C,P);
Opt.fista=true; Opt.gam=5e-2;
Opt.OutOp=OutputOpti(1,S'*gt,50);      
Opt.CvOp=TestCvgStepRelative(1e-4);  
Opt.maxiter=500; Opt.ItUpOut=50; 
Opt.run(S'*y);
        \end{lstlisting}
        };
        \node (FISTA_GR) at (0.5\textwidth,-4.9) {
        \begin{lstlisting}[linebackgroundcolor={%
                                \color{lightgrey}
                                \ifnum\value{lstnumber}=4 \color{orange!30} \fi
                                \ifnum\value{lstnumber}=8 \color{orange!30} \fi
                                \ifnum\value{lstnumber}=9 \color{orange!30} \fi
                                }]
%-- Regularizer ---------------
lamb=5e-3;
L=LinOpGrad(szin,[1 2]);
RL=CostGoodRoughness(L,1e-2);

%-- Optimization Algorithm --------
C = L2*S*H + lamb*RL;
Opt=OptiFBS(C,P);
Opt.fista=true; Opt.gam=5e-2;
Opt.OutOp=OutputOpti(1,S'*gt,50);      
Opt.CvOp=TestCvgStepRelative(1e-4);  
Opt.maxiter=500; Opt.ItUpOut=50; 
Opt.run(S'*y);
        \end{lstlisting}
        };
    \node[anchor=south] at (VMLMB_TVs.north) {Smoothed Total Variation \cite{aujol2009some,chambolle2010introduction}};
    \node[rotate=90,anchor=south] at (VMLMB_TVs.west) {VMLMB \cite{thiebaut2002optimization}};
    \node[anchor=south] at (VMLMB_GR.north) {Good's Roughness \cite{Verveer1999}};
    \node[rotate=90,anchor=south] at (FISTA_TVs.west) {FISTA \cite{beck2009fast}};
    \end{tikzpicture}
    \end{center}
    \caption{\globalbioim scripts for minimizing \eqref{eq:OptiPbDeconv} with differentiable regularizers $\mathcal{R}(\Ld \cdot)$. Differences with respect to the script corresponding to VMLMB with S-TV are highlighted.}
    \label{fig:ListingDiff}
\end{figure}

\subsection{Numerical Comparisons}

The modularity of \globalbioim, which was demonstrated in the scripts presented in Section~\ref{sec:deconv}, offers a simple way to compare the effect of regularizers as well as the efficiency of optimization algorithms.

\begin{figure}
    \centering
        \begin{tikzpicture}
    	\begin{groupplot}[group style={group size= 1 by 1},  
						  xmin=0.0001,xmax=0.1,
				   	      ymin=5,ymax=16,
						  title style={yshift=-0.10cm},
						  legend columns=2,
						  xmode=log,
						  grid=both,
						  legend pos= south west,   
					      legend style={legend cell align=left},
    					  width=0.7\textwidth,height=6.5cm]
    	\nextgroupplot[xlabel=$\lambda$,ylabel=SNR] 
    	 \addplot[blue,dashed,very thick] table{SNRvsLamb_TV.dat};
    	 \addplot[darkgreen,solid,very thick] table{SNRvsLamb_Hess.dat};
    	  \addplot[red,dashdotted,very thick] table{SNRvsLamb_TVs.dat};
    	   \addplot[orange,densely dotted,very thick] table{SNRvsLamb_GR.dat};
	    \legend{\footnotesize Total Variation, \footnotesize Hessian-Schatten, \footnotesize Smoothed-TV, \footnotesize Good's roughness};
		\end{groupplot}
		\end{tikzpicture}
    \caption{Evolution of the signal-to-noise ratio of the deconvolved image with respect to the regularization parameter $\lambda$.}
    \label{fig:SNR_vs_Lamb}
\end{figure}
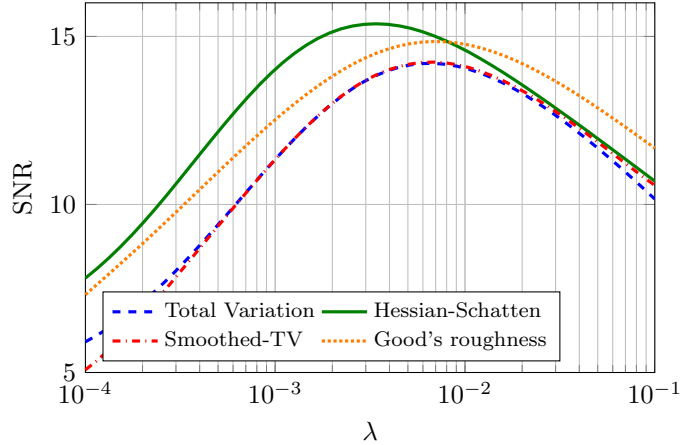

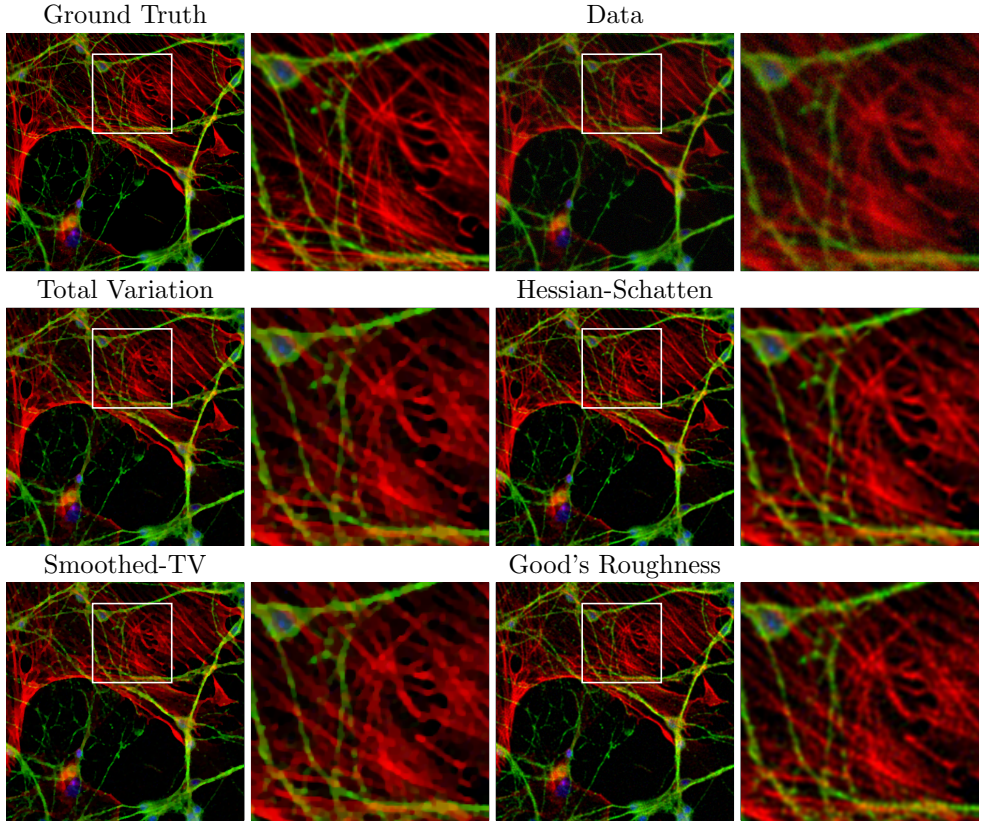
\begin{figure}
    \centering
    	\begin{tikzpicture}
		\begin{groupplot}[group style={group size= 4 by 3,                      
    					  horizontal sep=0.1cm,vertical sep=0.5cm}, 
						  xmin=0,xmax=460,
					   	  ymin=0,ymax=460,
						  title style={yshift=-0.2cm},
						  axis equal image,
						  grid style={black},
    					  width=0.42\textwidth]
    			    \nextgroupplot[enlargelimits=false,yticklabels={,,},xticklabels={,,},title=Ground Truth] 
    	 					 \addplot[] graphics[xmin=0,ymin=0,xmax=460,ymax=460] {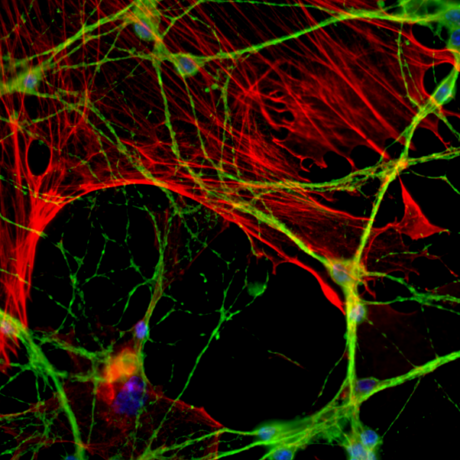};
    	 					 \draw[white,semithick]   (166.6,266.7) rectangle (320,420);
    	 			\nextgroupplot[enlargelimits=false,yticklabels={,,},xticklabels={,,}] 	
    	 					 \addplot[] graphics[xmin=-500,ymin=-800,xmax=880,ymax=580] {ground_truth.png};		 
    	 			\nextgroupplot[enlargelimits=false,yticklabels={,,},xticklabels={,,},title=Data] 	
    	 					 \addplot[] graphics[xmin=0,ymin=0,xmax=460,ymax=460] {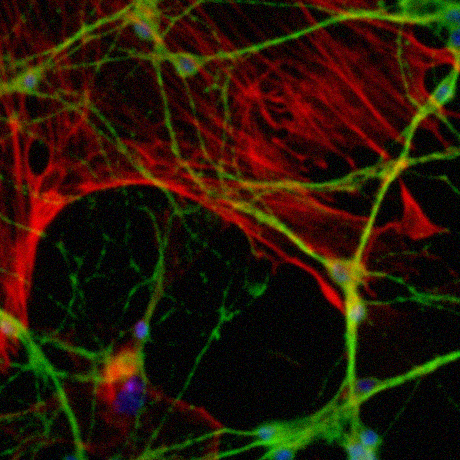};
    	 					 \draw[white,semithick]   (166.6,266.7) rectangle (320,420);
    	 			\nextgroupplot[enlargelimits=false,yticklabels={,,},xticklabels={,,}] 	
    	 					 \addplot[] graphics[xmin=-500,ymin=-800,xmax=880,ymax=580] {data.png};		 
    	 					 
    			   \nextgroupplot[enlargelimits=false,yticklabels={,,},xticklabels={,,},title=Total Variation] 	
    	 					 \addplot[] graphics[xmin=0,ymin=0,xmax=460,ymax=460] {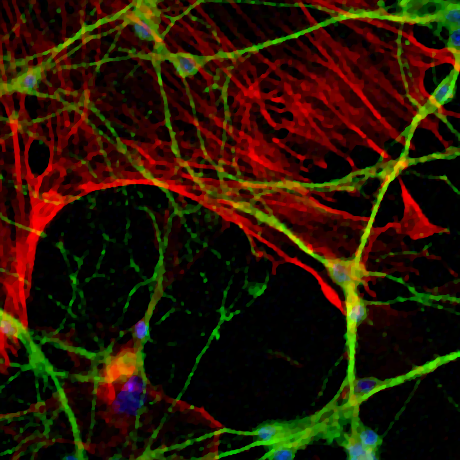};
    	 					 \draw[white,semithick]   (166.6,266.7) rectangle (320,420);
    	 		   \nextgroupplot[enlargelimits=false,yticklabels={,,},xticklabels={,,}] 	
    	 					 \addplot[] graphics[xmin=-500,ymin=-800,xmax=880,ymax=580] {DeconvBestSNR_TV.png};
    	 		   \nextgroupplot[enlargelimits=false,yticklabels={,,},xticklabels={,,},title=Hessian-Schatten] 	
    	 					 \addplot[] graphics[xmin=0,ymin=0,xmax=460,ymax=460] {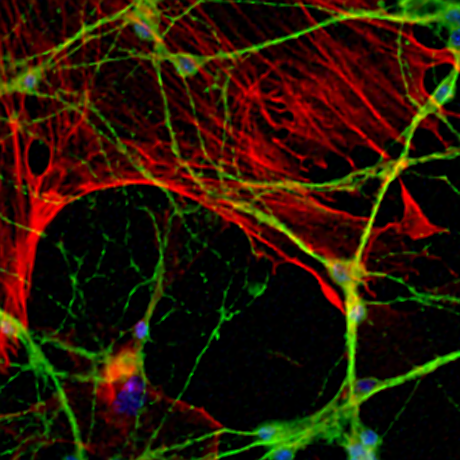};		
    	 					 \draw[white,semithick]   (166.6,266.7) rectangle (320,420);
    	 		   \nextgroupplot[enlargelimits=false,yticklabels={,,},xticklabels={,,}] 	
    	 					 \addplot[] graphics[xmin=-500,ymin=-800,xmax=880,ymax=580] {DeconvBestSNR_Hess.png};

    	 		    \nextgroupplot[enlargelimits=false,yticklabels={,,},xticklabels={,,},title=Smoothed-TV] 	
    	 					 \addplot[] graphics[xmin=0,ymin=0,xmax=460,ymax=460] {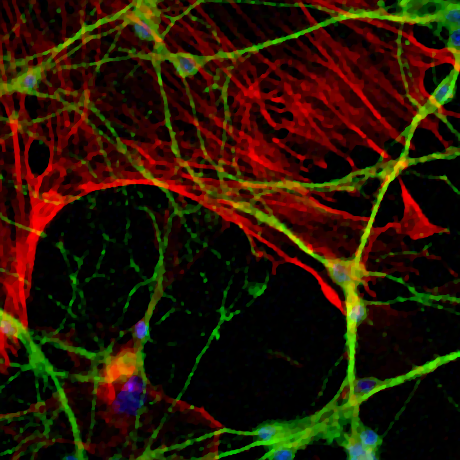};
    	 					 \draw[white,semithick]   (166.6,266.7) rectangle (320,420);
                    \nextgroupplot[enlargelimits=false,yticklabels={,,},xticklabels={,,}] 	
    	 					 \addplot[] graphics[xmin=-500,ymin=-800,xmax=880,ymax=580] {DeconvBestSNR_TVs.png};
    	 		    \nextgroupplot[enlargelimits=false,yticklabels={,,},xticklabels={,,},title=Good's Roughness,title style={yshift=-0.5ex}] 	
    	 					 \addplot[] graphics[xmin=0,ymin=0,xmax=460,ymax=460] {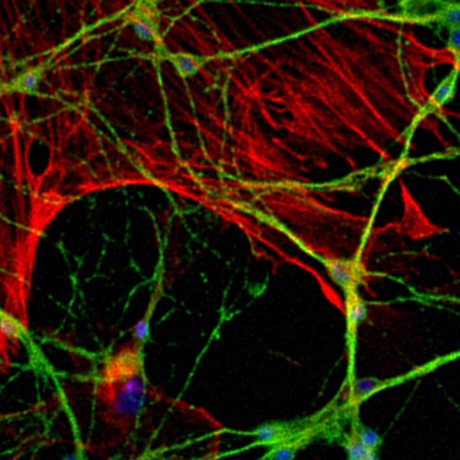};
    	 					 \draw[white,semithick]   (166.6,266.7) rectangle (320,420);
                    \nextgroupplot[enlargelimits=false,yticklabels={,,},xticklabels={,,}] 	
    	 					 \addplot[] graphics[xmin=-500,ymin=-800,xmax=880,ymax=580] {DeconvBestSNR_GR.png};
    	\end{groupplot}			
	\end{tikzpicture}
    \caption{Deconvolution results obtained with different regularizers for the optimal $\lambda$ extracted from Figure \ref{fig:SNR_vs_Lamb}. A zoom of the region delimited by the white square is also presented.}
    \label{fig:DeconvResults}
\end{figure}

We first present the quality of the deconvolution obtained with the four regularizers TV, HS, S-TV, and GR. Here, we used ADMM to minimize \eqref{eq:OptiPbDeconv} with non-differentiable regularizers (\ie TV and HS), and FISTA to minimize \eqref{eq:OptiPbDeconv} with differentiable regularizers (\ie S-TV and GR).
The SNR of the deconvolved image as a function of the regularization parameter $\lambda$ is depicted in Figure~\ref{fig:SNR_vs_Lamb}, while the deconvolved images that maximize the SNR are presented in Figure~\ref{fig:DeconvResults}. As expected, HS and GR lead to better results by avoiding the well-known staircasing effect of TV and S-TV.  Although GR is slightly below HS in terms of SNR, it provides comparable qualitative (\ie visual) results. 

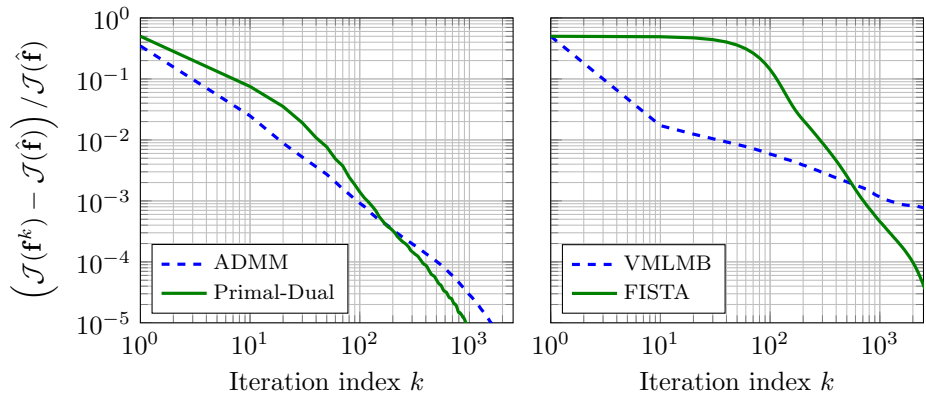
\begin{figure}
    \centering
       \begin{tikzpicture}
    	\begin{groupplot}[group style={group size= 2 by 1, horizontal sep=0.5cm},  
						  xmin=1,xmax=2500,
				   	      ymin=0.00001,ymax=1,
						  title style={yshift=-0.10cm},
						  xmode=log,
						  ymode=log,
						  grid=both,
						  legend pos= south west,   
					      legend style={legend cell align=left},
    					  width=0.5\textwidth]
    	\nextgroupplot[xlabel=Iteration index $k$,ylabel=$\left( \mathcal{J}(\fd^k) - \mathcal{J}(\hat{\fd})\right)/\mathcal{J}(\hat{\fd})$]
    	 \addplot[blue,dashed,very thick] table{CostvsIters_RelativeDiff_ADMM.dat};
    	 \addplot[darkgreen,solid,very thick] table{CostvsIters_RelativeDiff_PD.dat};
    	 \legend{\footnotesize ADMM, \footnotesize Primal-Dual};
    	\nextgroupplot[xlabel=Iteration index $k$,yticklabels={,,}] 
    	    	 \addplot[blue,dashed,very thick] table{CostvsIters_RelativeDiff_VMLMB.dat};
    	 \addplot[darkgreen,solid,very thick] table{CostvsIters_RelativeDiff_FISTA.dat};
    	 \legend{\footnotesize VMLMB, \footnotesize FISTA};
    \end{groupplot}
    \end{tikzpicture}
    \caption{Convergence curves for the minimization of \eqref{eq:OptiPbDeconv} with TV (left) or S-TV (right). The solution $\hat{\fd}$ has been computed by performing 10,000 iterations of ADMM (FISTA, respectively).}
    \label{fig:CvgCurves}
\end{figure}

 We now fix the parameter $\lambda$ to the value that maximizes  the SNR  in Figure \ref{fig:SNR_vs_Lamb} for TV and S-TV. The convergence curves generated by ADMM and the primal-dual method for the minimization of \eqref{eq:OptiPbDeconv} with TV, as well as those generated by FISTA and VMLMB when the regularizer is set to be S-TV, are presented in Figure \ref{fig:CvgCurves}.
 We would like to emphasize that the parameters of the  algorithms have not been tuned to obtain the fastest convergence. Hence, these results constitute more an illustration of the kind of comparisons that can be easily performed with \globalbioim rather than an empirical demonstration of the convergence speed of these algorithms. Moreover, both ADMM and the primal-dual method offer alternative splitting strategies that may lead to improved convergence speed. Note that the adaptation of the scripts in Figure \ref{fig:ListingNonDiff} to these variations  is straightforward with \globalbioim. We refer the reader to the online documentation of the corresponding two \opti classes for more details on how to establish such adaptations.

\section{Discussion}

Open-source software is an essential component of modern research. Not only does it shape theoretical developments, but it also turns out to be a critical tool to bridge the gap that separates researchers specialized  in computer science/mathematics from scientists versed in biophysical sciences/medicine. Moreover, open-source software can act as a catalyst for engaging in new collaborations by promoting external contributions.

Motivated by the observation that the image-formation models of most of the commonly used biomedical imaging systems  can be expressed as a composition of a limited number of elementary operators,  we developed the open-source MATLAB library \globalbioim. 
This library provides a unified and user-friendly framework for the resolution of inverse problems. 
It is designed around three entities, namely, forward models, cost functions, and optimization algorithms, which constitute the building blocks of any inverse problem. This organization gives \globalbioim a modularity that greatly facilitates the comparison between  regularizers and or solvers, as illustrated in Section \ref{sec:appli}.
Moreover, \globalbioim enjoys an operator-algebra mechanism able to perform automatic simplification of composed operators.  
Finally, new modalities, cost functions, or solvers are easily added to the framework of  \globalbioim. Reference papers that use \globalbioim is provided in the Section ``Related Papers'' of the online documentation \url{http://bigwww.epfl.ch/algorithms/globalbioim/}.

\section*{Acknowledgements}

The authors would like to thank warmly Rainer Heintzmann for fruitful discussions related to this project as well as Philippe Th\'evenaz for his useful feedback on the paper.
This research was supported  by the European Research Council (ERC) under the European Union’s Horizon 2020 research and innovation programme, Grant Agreement No.\ 692726 GlobalBioIm: Global integrative framework for computational bio-imaging.

\section*{References}
\bibliographystyle{plain}
\bibliography{refs.bib}

\end{document}